# Early Evolution of Earth-Directed Coronal Mass Ejections in the Vicinity of Coronal Holes

Suresh Karuppiah[1] · Mateja Dumbović[1] · Karmen Martinić[1] ·
Manuela Temmer[2] · Stephan G. Heinemann[3,4] · Bojan Vršnak[1]

[1] Hvar Observatory, Faculty of Geodesy, University of Zagreb, Croatia
[2] Institute of Physics, University of Graz, Austria
[3] Department of Physics, University of Helsinki, P.O. Box 64, 00014, Helsinki, Finland
[4] Max-Planck-Institut für Sonnensystemforschung, Justus-von-Liebig-Weg 3, 37077 Göttingen, Germany

## Abstract

We investigate the deflection and rotation behavior of 49 Earth-directed coronal mass ejections (CMEs) spanning the period from 2010 to 2020 aiming to understand the potential influence of coronal holes (CHs) on their trajectories. Our analysis incorporates data from coronagraphic observations captured from multiple vantage points, as well as extreme ultraviolet (EUV) observations utilised to identify associated coronal signatures such as solar flares and filament eruptions. For each CME, we perform a 3D reconstruction using the Graduated Cylindrical Shell (GCS) model. We perform the GCS reconstruction in multiple time steps, from the time at which the CME enters the field of view (FOV) of the coronagraphs to the time it exits. We analyse the difference in the longitude, latitude, and inclination between the first and last GCS reconstructions as possible signatures of deflection/rotation. Furthermore, we examine the presence of nearby CHs at the time of eruption and employ the Collection of Analysis Tools for Coronal Holes (CATCH) to estimate relevant CH parameters, including magnetic field strength, centre of mass, and area. To assess the potential influence of CHs on the deflection and rotation of CMEs, we calculate the Coronal Hole Influence Parameter (CHIP) for each event and analyse its relationship with their trajectories. A statistically significant difference is observed between CHIP force and the overall change in a CME's direction in the lower corona. The overall change in a CME's direction accounts cumulatively for the change in latitude, longitude, and rotation. This suggests that the CHIP force in the low corona has a significant influence on the overall change in the direction of Earth-directed CMEs. However, as the CME evolves outward, the CHIP force becomes less effective in causing deflection or rotation at greater distances. Additionally, we observe a negative correlation between the deflection rate of the CMEs and their velocity, suggesting that higher velocities are associated with lower deflection rates. Hence, the velocity of a CME, along with the magnetic-field from CHs, appears to play a significant role in the deflection of CMEs. By conducting this comprehensive analysis, we aim to enhance our understanding of the complex interplay between CHs, CME trajectories, and relevant factors such as velocity and magnetic field strength.

## 1. Introduction

Coronal mass ejections (CMEs) are the major eruptive phenomena in the solar atmosphere. They cause various space-weather effects and can trigger severe geomagnetic storms if their arrival on Earth is accompanied by a certain magnetic configuration (Temmer, 2021). Therefore, it is very important to understand the effects that might impact their evolution. Previous studies have shown that CMEs in general may exhibit non-radial propagation, i.e. be deflected/rotated. For instance, Cremades and Bothmer (2004) found that CMEs are deflected toward low latitudes during solar minimum. Shen et al. (2011) found that the deflection could be due to the non-uniform distribution of the magnetic fields and tends to be in the direction of the lower background magnetic-energy density. Gui et al. (2011) found a positive correlation of the deflection rate with the strength of the magnetic-energy density and a negative correlation between the deflection rate and the CME velocity. Kay et al. (2017) found that the effective deflections occur below $2R_\odot$ and vary from $8.9^o$ to $26.7^o$ in latitude and $0.2^o$ to $10^o$ in longitude. They also suggested that the deflection in latitude is due to the coherent force, which deflects the CME towards the heliospheric current sheet (HCS), and that the deflection in longitude results from the small-scale magnetic gradient of active regions.

Isavnin, Vourlidas, and Kilpua (2013) investigated 15 CMEs during the solar minimum period of 2008 to 2010 from the Sun to 1 AU and observed longitudinal deflections within a few solar radii from the Sun, while the orientation of the CMEs continued to change up to 1 AU. In a subsequent study by savnin, Vourlidas, and Kilpua (2014), they reported that deflection and rotation were still observed below $30R_\odot$, but significant deflection and rotation were further observed between $30R_\odot$ and 1 AU. Kay, Opher, and Evans (2015) analysed CME deflections due to magnetic forces using a three-dimensional version of ForeCAT model (Kay, Opher, and Evans, 2013) that simulates CME evolution in the corona. They observed that the majority of deflections occurred below $10R_\odot$, suggesting that both global and local magnetic-field gradients contribute to the total deflection. In another study by Kay and Opher (2015), it was found that the magnetic forces are responsible for effective deflection and rotation below 2 $R_\odot$, with 1% of CMEs experiencing a deflection from 5 $R_\odot$ to 1 AU. Beyond 10 $R_\odot$, they observed that 10% of rotations. Wang et al. (2014) examined a single CME event that occurred on 12 September 2008 and confirmed the deflected propagation in both low corona and in interplanetary space, suggesting that the majority of deflection occurs in the interplanetary medium influenced by the solar wind and interplanetary magnetic field. Wang et al. (2016) further studied a CME event 15 March 2015 that resulted in a large geomagnetic storm. The study suggested that the CME underwent a $12^o$ eastward deflection in the interplanetary medium before reaching Earth although it had a westward orientation in the LASCO field of view (FOV), making it a geoeffective event.

Vourlidas et al. (2011) investigated the rotational behaviour of a CME event in the quiet Sun on 10 June 2010, and suggested that CME rotations may be due to the disruption of one of the flux-rope foot points at the beginning of the eruption and not related to the Lorentz force. In the study of Capannolo et al. (2017), the *cartwheel* shaped CME that occurred on 09 April 2008, was investigated. They observed a distinctive deflection and rotation of this particular CME, which was attributed to the asymmetric reconnection process during its eruption. Interestingly, the study revealed that this unique behavior was not influenced by magnetic forces in the surrounding background. Yurchyshyn, Abramenko, and Tripathi (2009) analysed the angle between the orientation of CMEs and post-eruption arcades (PEAs) for 100 events and found that the majority of the CMEs lie in the direction of the axial fields of the PEAs with an orientation difference of $10^o$ for most of the events in the direction of the solar Equator and HCS. In an another study, Kliem, Török, and Thompson (2012) found both the force of an external shear-field component and the relaxation of twist components are potentially very significant contributors to the rotation. They suggested further that the rotation due to twist relaxation tends to act mainly low in the corona, in a height range up to only a few times the distance between the foot points of the erupting flux, whereas the rotation by the shear field tends to be distributed across a larger height range. The magnetic reconnection contributes weakly to the rotation.

Coronal holes (CHs) refer to areas on the Sun's surface exhibiting reduced density and lower temperature, as well as diminished magnetic-field intensity, while displaying higher velocities of plasma outflow compared to the surrounding regions. At solar minimum, the Sun's magnetic field is primarily characterised by a rotationally aligned dipole component, leading to the formation of extensive coronal holes enveloping the north and south polar caps of the Sun. During solar maximum, coronal holes have the potential to emerge at various latitudes across the solar surface. However, their duration is relatively short-lived, lasting only a few solar rotations before undergoing transformations into diverse magnetic configurations (Cranmer, 2009). Several authors have been working on the properties of CH. For example, Heinemann et al. (2018a) conducted a study that focused on a specific CH event that occurred in 2012, approximately spanning ten solar rotations. They investigated how the three distinct evolutionary stages of the CH, as observed in the solar atmosphere, influenced the properties of the associated high-speed streams at 1 AU. Heinemann et al. (2018b) continued their investigation of the same CH and concluded that the small-scale structures of strong unipolar magnetic fields are the fundamental building blocks of a CH. In the study conducted by Hofmeister et al. (2019), a group of CHs was examined to analyse their photospheric magnetic structure. The research revealed a strong correlation between the number of magnetic bright points within magnetic elements and the area of those elements in the CHs and also found that the total area covered by long-lived magnetic elements determines the unbalanced magnetic flux of the Chs.

When a CME approaches a CH, the open magnetic configuration of the CH acts as a magnetic barrier, leading to a change in the trajectory of the CME. More specifically, it was found that CHs may also cause deflection/rotation of CMEs. Kilpua et al. (2009) investigated two high-latitude CMEs and reported the equatorward deflection of these CMEs due to polar coronal holes during solar minimum. Gopalswamy et al. (2004) suggested that the deflection of CMEs may be caused by the open magnetic field from the nearby CH, and that this deflection may be toward the Sun-Earth line or away from it, depending on the relative positions of the CHs. In a follow-up study, opalswamy et al. (2009b) reported the first detection of the deflection of an extreme-ultraviolet (EUV) wave associated with a CME by the CH. Gopalswamy et al. (2009a) investigated how the presence of nearby CHs affects the trajectory of CMEs. They introduced the Coronal Hole Influence Parameter (CHIP, see Equation eqn1.), which is a force calculated from CH properties, for a set of CMEs associated with driverless interplanetary shocks in Solar Cycle 23, and found that the CHIP values of ejecta whose magnetic structure was deflected from the Sun-observer line, were twice those of magnetic clouds (MCs) where the flux-rope magnetic structure was clearly observed at the Sun-observer line.

In a study conducted by Mohamed et al. (2012), a total of 29 CMEs during Solar Cycle 23 were investigated, and their associated CHIP values were analysed. The findings confirmed that CMEs in close proximity to CHs experience the strongest influence from these CHs. Such CMEs often exhibit driverless shocks due to the deflection of their magnetic structure caused by the CHs. Another study by Mäkelä et al. (2013) examined 54 CMEs and provided further insights. It was revealed that all of the analysed CMEs possess a magnetic-flux-rope structure. Notably, for certain CMEs with a CHIP value exceeding 2.6 G, the influence from nearby CHs results in deflection of the CME trajectory away from the Sun-observer line. Consequently, these CMEs are observed as non-magnetic clouds (non-MCs). CHs display a prevailing magnetic-field orientation, leading to an open magnetic-field configuration. Along these magnetic lines, solar plasma is propelled outward into interplanetary space, giving rise to what is known as high-velocity solar-wind streams (HSS: Nitti et al., 2023, and references therein). As per the findings of Gopalswamy et al. (2022), it has been proposed that the high intensity of the geomagnetic storm may be attributed to the prolonged duration of the southward magnetic-field alignment of the CME occurring on 20 August 2018. This effect could potentially have been amplified by the subsequent HSS-induced compression.

Palmerio et al. (2022) investigated the same event and proposed that the CME underwent additional deflection and rotation as it traversed the region between Earth and Mars' orbits, primarily influenced by its interaction with a subsequent HSS originating from a CH. In another investigation, ahade, Cécere, and Krause (2020) employed magnetohydrodynamic (MHD) simulations to study the evolution of CMEs in the vicinity of CHs. Their findings indicated that the degree of deflection experienced by CMEs was more substantial when encountering wider CHs with stronger magnetic fields. Conversely, as the CME moved away from the CH, the level of deflection decreased. In a study conducted by Bosman et al. (2012), the three-dimensional characteristics of 51 CMEs occurring between 2007 and 2010 were investigated using a forward modeling technique. The findings revealed that approximately 82% of the examined events exhibited displacement from their original source position, migrating towards lower latitudes. This observation provided confirmation of the deflection of CMEs towards the solar Equator. Heinemann et al. (2019a) examined a specific CME that occurred on 21 June 2011, and investigated its interaction with HSS originating from CHs. They discovered that this interaction commences at a height ranging from 1.3 $R_\odot$ to 3 $R_\odot$, resulting in a significant deflection of the CME by approximately 30°.

In this study, we systematically analyse possible deflections/rotations of CMEs in relation to CHs in the low, middle, and outer corona using low coronal signatures and 3D reconstruction of CMEs. Our aim is to confirm via a statistical, observational study that CHs indeed may cause deflection/rotation and to analyse at which distances we can expect this to occur. Section 2 explains the selection of data and the methods used. A detailed analysis of GCS reconstruction is presented in Section 2.1, whereas an estimation of CH parameters is presented in Section 2.2. Finally, we present and discuss our results in Section 3.

## 2. Data and method:

We collect 49 CME-ICME pairs for this study from the ICME catalog (www.srl.caltech.edu/ACE/ASC/DATA/level3/icmetable2.htm) of Richardson and Cane (2010) in the time period 2010 to 2020. The time period for the analysis is based on the availability of data used for that study. This includes stereoscopic data obtained using the *Sun Earth Connection Coronal and Heliospheric Investigation (SECCHI: Howard et al., 2008)* suite onboard the *Solar TErrestrial RElations Observatory* spacecraft (STEREO: Kaiser et al., 2008) and the C2/C3 coronagraphs aboard the *Solar and Heliospheric Observatory* spacecraft (SOHO: Domingo, Fleck, and Poland, 1995). In particular, the SECCHI instrument captures images through both the inner corona (COR1), covering a range of 1.1 to 3 $R_\odot$, and the outer corona (COR2), which extends from 2 to 15 $R_\odot$. We use white-light coronagraph data from at least two different vantage points in order to perform a 3D reconstruction of the observed CME. For that purpose, we apply the Graduated Cylindrical Shell (GCS) model of Thernisien (2011). The GCS model has a wire frame resembling a croissant shape, designed to mimic the morphology of CME flux ropes (Vourlidas et al., 2013) on images obtained nearly simultaneously from multiple views: e.g. from SOHO and STEREO. The leading edge of the CME can be tracked up to 25 $R_\odot$. From this model, we derive a 3D height and speed that should represent the true height and speed within the limitations of the idealised model assumptions. A number of researchers have used this GCS model in case of analysing the Earth-directed CMEs (Gui et al., 2011; Temmer et al., 2017, 2021; Suresh, Gopalswamy, and Shanmugaraju, 2022). Also, we consider the time period of data availability of the *Atmospheric Imaging Assembly* (AIA: Lemen et al., 2012) EUV telescope onboard the *Solar Dynamics Observatory (SDO: Pesnell, Thompson, and Chamberlin, 2012)* to observe and analyse CME low coronal signatures (Hudson and Cliver, 2001) and CHs at high resolution. For the period 2010 to 2020, there are initially 94 events in the Richardson and Cane catalogue. After eliminating some events based on the following criteria, we are left with 49 CMEs: i) There are 9 events before the SDO data period ii) 15 events have signatures too faint to perform GCS iii) 12 events have no observations from at least 2 vantage points iv) 2 events are under suspicion that their source regions were situated on the Sun's far side, and this is consistent with the information provided in the DONKI catalogue. v) Additionally, GCS reconstruction for 5 events could not be reliably performed due to CME-CME interactions in the corona. vi) one event is not found in the LASCO catalog at the given time and we cannot establish a reliable connection to a specific CME vii) for another event, reliable GCS reconstruction is not possible due to data calibration issues.

## 2.1 Deriving CME Deflection/Rotation

For each of the 49 events, we analyse the low coronal signatures (LCS) such as solar flares, post-eruption arcades (PEAs), dimmings, and eruptive prominences using SDO/AIA 193/131 Å images to determine the source region. This includes 3 stealth CMEs, which had no clear eruptive signatures (Nitta and Mulligan, 2017; Nitta et al., 2021), however small brightenings and/or movement of coronal loops indicated their source region. Whenever possible, the locations are cross-checked with the SOHO/LASCO halo CME catalogue (Gopalswamy et al., 2010, : cdaw.gsfc.nasa.gov/CME_list/halo/halo.html) and the DONKI catalogue (kauai.ccmc.gsfc.nasa.gov/DONKI/search/).

We next perform GCS reconstructions of the CMEs in multiple time steps as they cross the FOV of the COR1 and COR2 cameras onboard STEREO. The GCS model, by considering CMEs as idealised flux-rope structure, is applied to the white-light images from STEREO and LASCO at the closest time of each other. The GCS fit has six independent parameters: i) latitude and ii) longitude of the propagation direction, iii) height of the CME's apex, iv) half-angular width of the shell, v) tilt angle of the flux-rope central-axis orientation, and vi) aspect ratio: the ratio between major and minor radii of the flux rope. In performing the GCS reconstruction, we treat longitude, latitude, inclination, and height as free parameters, while aspect ratio and half angle remain fixed since this model assumes self-similar expansion of the flux rope. Some events are tracked only in COR2 due to their faint signatures in COR1 FOV. For each CME, we analyse whether the longitude, latitude, and/or inclination changed between the first and the last reconstruction. The difference between the longitude/latitude/tilt measured at the first and the last reconstruction indicates possible deflection/rotation of the CME. We also calculate possible deflection in the low corona by looking at the differences in the longitudes/latitudes between the source region determined from low coronal signatures and the apex position obtained from the first GCS reconstruction. The results for all of the events are presented in Table 2 (see appendix). The date and time of the first C2 appearance are given in the first and second columns. Stealth CMEs are marked with an asterisk symbol in the first column. The source location of each event and the type of low coronal signature is given in columns 3 and 4. Column 5 contains a number of time steps performed by the GCS reconstruction. The longitudinal (Δlon) and latitudinal (Δlat) deflection, and the rotation (Δrot) of each event is given in columns 6 to 8. The number of CHs associated with each CME is listed in column 8. The 3D speed of each CME, calculated using linear regression based on multiple height-time measurements is given in the last column. The detailed list of GCS reconstructions and the calculated CHIP values of all events are available at figshare.com/s/b1b162e3a28da5d29a30.

Figure 1 shows one of the GCS reconstructed events. This CME was observed on 12 July 2012 with LASCO-C2 first appearance time 16:48 UT. The source location is W03S17, inferred from the low coronal signature of an X-class flare in the SDO/AIA 193/131 Å image (not shown). The STEREO-A and -B (hereinafter STA and STB) spacecraft are located at 120$^o$ West and 115$^o$ East from the Sun-Earth line, respectively. Since this CME propagated 3$^o$ away from the Sun-Earth line towards the West, the STA and STB spacecraft observed this CME at a distance of 117$^o$ East and 118$^o$ West from the CME direction, respectively. We perform a GCS reconstruction of the CME at 12 times from 16:25 UT (COR1) to 18:39 UT (COR2). The results are shown in Table 1, which shows that the longitude is the same at all times, indicating there is no visible deflection in the longitude. The latitude varies from -18$^o$ to -11$^o$, indicating a deflection of 7$^o$ toward the North. In addition, the tilt angle varies from 0 to 72$^o$, indicating a rotation of 72$^o$ from the solar Equator in the counterclockwise direction.

Kay and Gopalswamy (2018) analysed the same event by tracking the radial expansion and evolution of the CME using the ForeCAT model out to the distance 20 $R_\odot$ and find changes in longitude, latitude, and tilt by -1.9, -0.4, and 1.7$^o$, respectively. Hinterreiter et al. (2021) reported the GCS parameters for the event on 12 July 2012 as longitude 8$^{o}$, latitude -12$^{o}$, and tilt of 68$^{o}$ within the COR2 FOV by considering the time of the image when the flux-rope is seen clearly in all viewpoints. Note that the value of our tilt is closer to that derived by Hinterreiter et al. (2021).

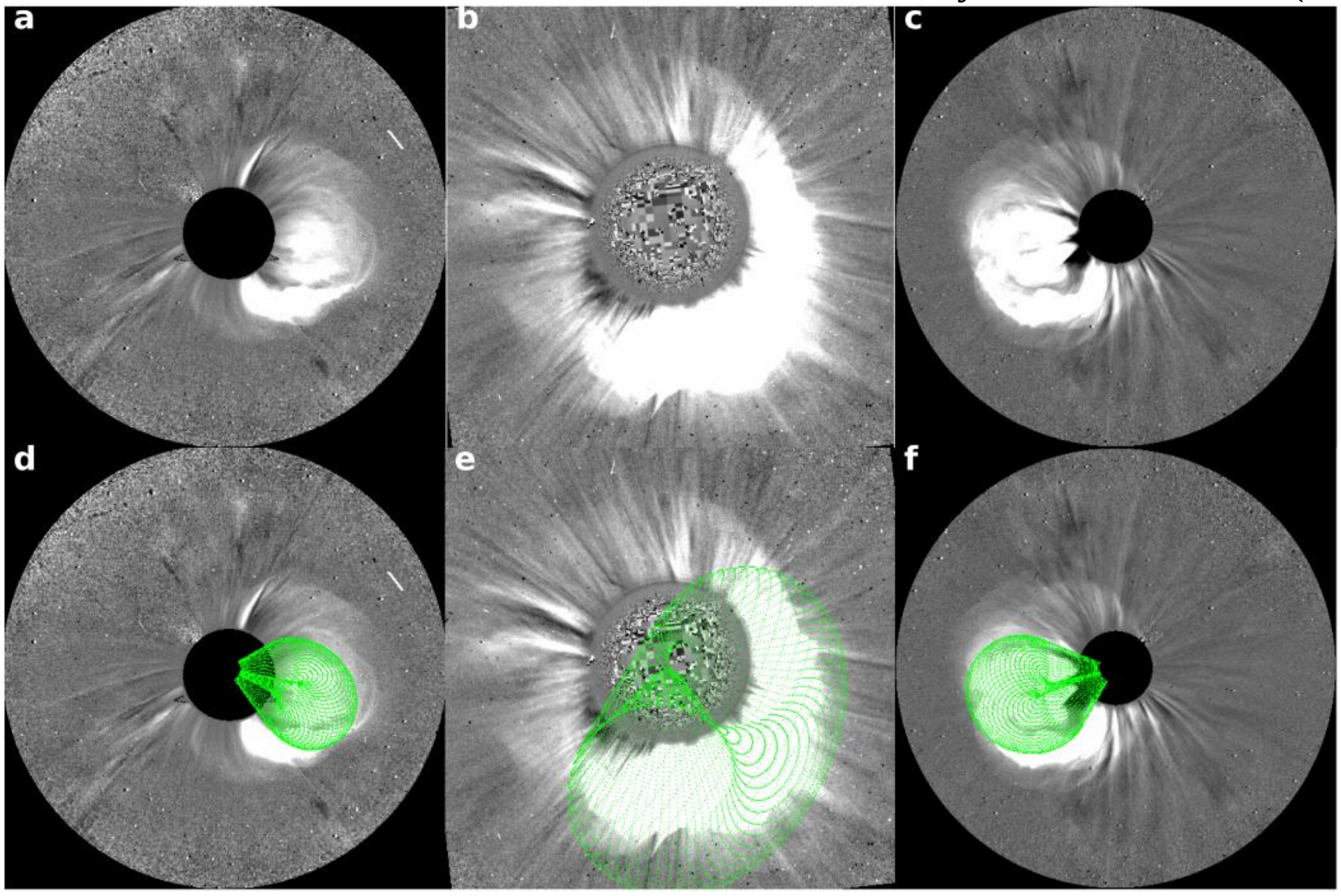

Figure 1. Multiple observations of the CME on 12 July 2012. (a) and (d) show STEREO-A, whereas (c) and (f ) show STEREO-B. (b) and (e) show LASCO-C2. The green mesh in bottom panels shows the GCS reconstruction.

| Date | Detector | Time | Longitude | Latitude | Tilt | Height | Ratio | H.angle |
|---|---|---|---|---|---|---|---|---|
| 12 July 2012 | COR1 | 1625 | 8 | -18 | 0 | 1.9 | 0.35 | 20 |
| " | " | 1630 | 8 | -18 | 0 | 2.1 | 0.35 | 20 |
| " | " | 1635 | 8 | -18 | 6 | 2.42 | 0.35 | 20 |
| " | " | 1640 | 8 | -18 | 7 | 2.8 | 0.35 | 20 |
| " | " | 1645 | 8 | -18 | 20 | 3 | 0.35 | 20 |
| " | " | 1650 | 8 | -18 | 27 | 3.42 | 0.35 | 20 |
| " | " | 1655 | 8 | -18 | 31 | 3.93 | 0.35 | 20 |
| " | COR2 | 1724 | 8 | -11 | 31 | 7.57 | 0.35 | 20 |
| " | " | 1739 | 8 | -11 | 46 | 9.4 | 0.35 | 20 |
| " | " | 1754 | 8 | -11 | 60 | 11.1 | 0.35 | 20 |
| " | " | 1824 | 8 | -11 | 72 | 14.42 | 0.35 | 20 |
| " | " | 1839 | 8 | -11 | 72 | 16.3 | 0.35 | 20 |

Table 1: GCS reconstruction results for the CME observed at LASCO-C2 on 12 July 2012.

Out of the 49 CMEs, we find changes between the first and last step of the GCS reconstruction in the longitude for 7 CMEs (14%), in latitude for 23 CMEs (47%), and for tilt in 19 CMEs (39%). However, we note that not every change in longitude, latitude, and inclination can be considered a deflection and rotation. Verbeke et al. (2023) have shown that the GCS parameters may change from one observer to another (i.e. from one observation to another) resulting on average in differences in longitude of 11$^{o}$, in latitude of 6$^{o}$, and in inclination of 25$^{o}$ for the same reconstruction. Therefore, we consider as significant only those changes in longitude, latitude, and rotation values that are larger than those. According to these criteria, only 1 event shows deflection in longitude, 10 in latitude, and 11 in rotation (i.e. change in the tilt).

We note that there are 5 events that show sudden variation in the tilt between GCS reconstructions performed on COR1 and COR2 images. However, this seems not to be related to the change of the observing instrument. Figure 2 shows one such event (CME on 18 January 2012). The top row shows running difference images of STEREO-A and -B and the corresponding GCS fit is given in the lower panels. The first and last reconstructions in COR1 are at 12:00 and 13:30 UT, respectively, and their inclinations are $15^{o}$ and $29^{o}$, whereas the first and last reconstructions in COR2 are at 15:54 and 16:54 UT, and their inclinations are both $72^{o}$. There is no gradual increase or decrease in the inclination between the last reconstruction of COR1 and the first reconstruction of COR2 time frames. This may be due to the larger time difference between the last reconstruction of COR1 and the first reconstruction of COR2. The presence of a larger time difference is due to unclear flux-rope signatures after the last reconstruction in COR1 and first reconstruction in COR2. Note that the size of the occulting disk is not symmetric in COR2-A and -B because their relative position with respect to the Earth is not identical.

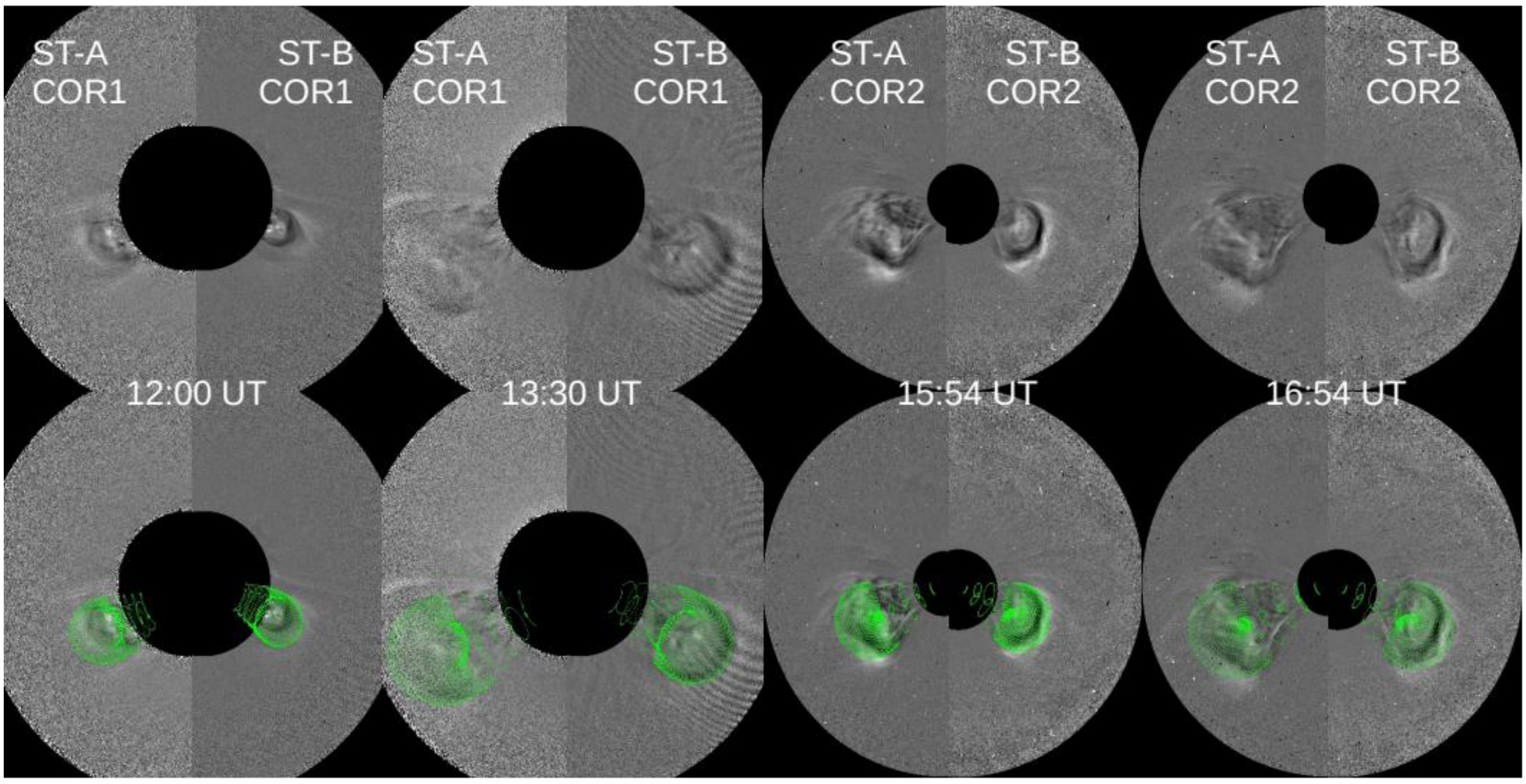

Figure 2. GCS reconstruction of the CME observed on 18 January 2012 from COR1 to COR2. *The top panels* show running difference images in COR1 and COR2, whereas GCS reconstructions are provided in *the bottom images (green mesh)*. Note that only half-disc images are given for STEREO -A and -B (*as marked in the figure)*. Four different time-steps correspond to the first and last reconstructions in COR1 and COR2. Note that the tilt derived from the last reconstruction in COR1 (1330UT) is $29^{o}$ and changes to $72^{o}$ in the first reconstruction in COR2 (1554UT).

## 2.2 Deriving Coronal-Hole Properties

We check for the presence of CHs on the solar disc on the day of the eruption by analysing SDO/AIA 193 Å images and the Boulder catalogue (www.ngdc.noaa.gov/stp/space-weather/solar-data/solar-imagery/composites/full-sun-drawings/boulder). In addition, we consult CH identification results performed by the Coronal Hole Identification via Multi-thermal Emission Recognition Algorithm (CHIMERA; Garton, Gallagher, and Murray, 2018), available on SolarMonitor (www.solarmonitor.org) and Spatial Possibilistic Clustering Algorithm (SPOCA: Verbeeck et al., 2014) in JHelioviewer (Müller et al., 2017). We use the Collection of Analysis

Tools for Coronal Holes (CATCH; Heinemann et al., 2019b) on SDO/AIA 193 Å filtergrams to extract the boundary of each CH. This is done by applying the optimum threshold that is derived from the intensity gradient across the CH boundary. Using the extracted boundary, CATCH automatically calculates the CHs morphological (area and geometric centre of mass) properties as well as the properties of the underlying photospheric magnetic field. The magnetic field is calculated from *Helioseismic and Magnetic Imager* (HMI) 720 s LoS magnetograms onboard SDO (HMI: Scherrer et al., 2012).

Next, we calculate the Coronal Hole Influence Parameter (CHIP), which is a measure of a force (fictitious force; Cremades, Bothmer, and Tripathi, 2006) acting from the CHs on the CME source region, proportional to the product of the magnetic field and the area of the CH, and inversely proportional to the square of the distance from the source to the CH (see Gopalswamy et al., 2009a, and references therein). CHIP is given by:

$$CHIP = \frac{B.A}{d^2}\,\hat{e}\ [G]$$

where $B$ is the line-of-sight magnetic-field strength, $A$ is the CH area and $d$ is the distance between the source location of the CME (i.e. the projection of its apex on the solar disc) and the centre of mass of the CH. $\hat{e}$ is a unit vector pointing from the CH to the CME source region. In case of more than one CH present at the disc, we calculate the CHIP value of each CH separately and take the vector sum of all CHIP values as the total CHIP of that associated event.

Figure 3 shows CHs on the AIA-193 Å image (green on left) and the corresponding HMI magnetogram (red on right) for the 12 July 2012 event. The source region is indicated by ``X" symbol on both AIA-193 Å image. The properties of two CHs are listed in Table 3. The source location of this CME as estimated from low coronal signatures is S17W03. Two CHs were observed for this event located at S38E32 and at N25E34. The area and the signed mean magnetic-field strength [$B$] of CHs are derived from CATCH. The calculated distance between the source region and the centre of mass of the first and second CH is 3.82 x $10^5$ km and 4.13 x $10^5$ km, respectively. The source region is determined using **JHelioviewer**, ensuring the accuracy of the calculated distances. The corresponding CHIP values, using Equation 1, are 0.53 G and 0.26 G, respectively. The total value of CHIP is 0.46 G. This means that the CME is influenced by the nearby CHs with a CHIP of 0.46 G. The properties of all CHs associated with each CME are listed in Table 1 provided at figshare.com/s/b1b162e3a28da5d29a30.

The primary parameter for determining the CHIP value of CHs is the distance between the location of CME and the CH. A larger distance indicates a less dominant influence over the associated CMEs. This influence is quantified through three distinct CHIP values, each corresponding to different distance ranges. CHIP1 represents the distance between the source position and the CH, indicating the CH's influence at the eruption's onset. CHIP2 calculates the distance between the CH and the location on the solar disk which corresponds to the projection of the CME's apex at the time of the first GCS reconstruction, reflecting the CH's effect as the CME enters the COR1 instrument's field of view. CHIP3 measures the distance between the CH and the location on the solar disk which corresponds to the projection of the CME's apex at the time of the last GCS reconstruction, capturing the CH's influence when the CME reaches its maximum distance within the COR2 field of view. These CHIP values allow us to comprehensively assess the varying impact of CHs on CMEs at different stages.

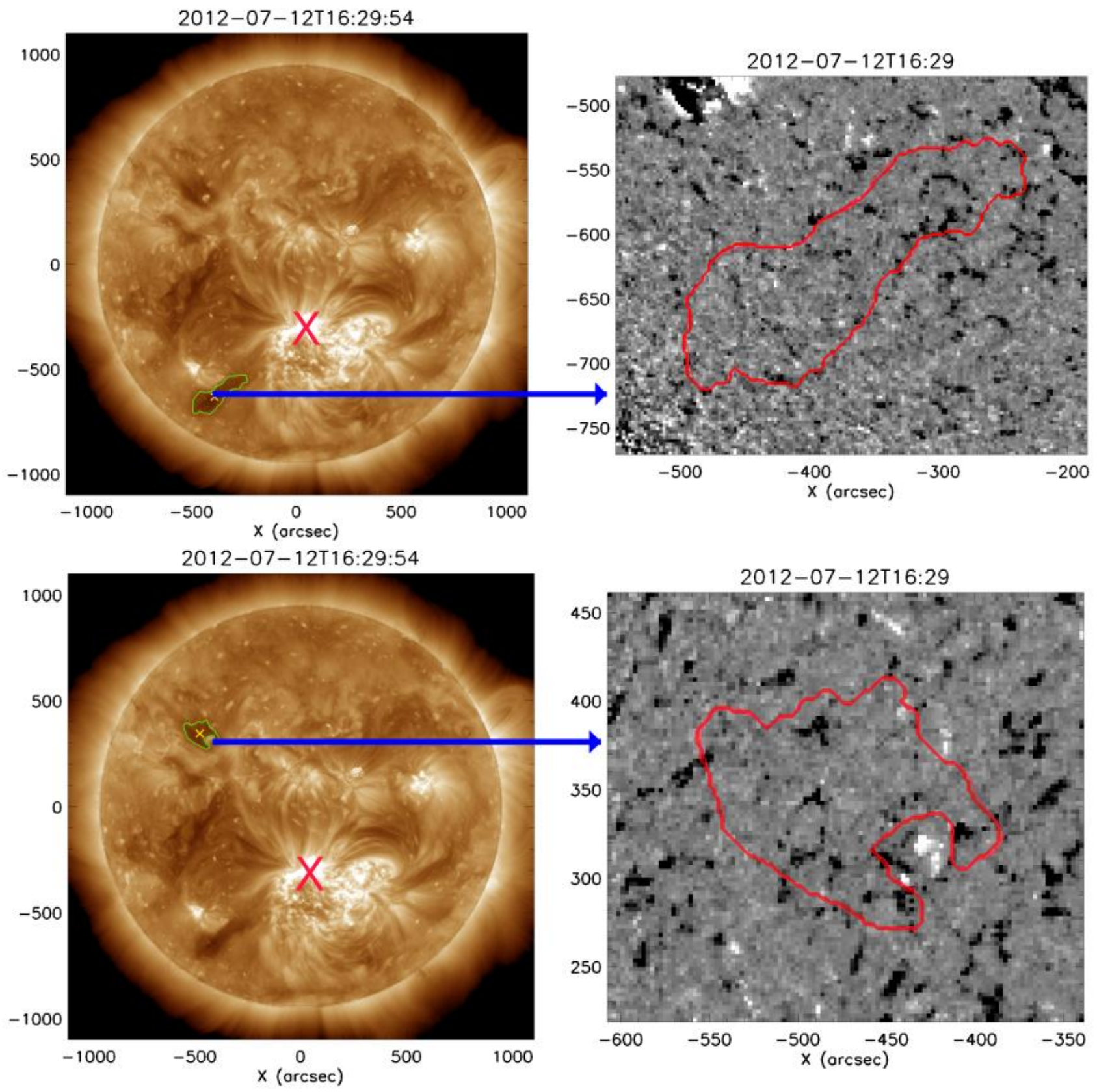

Figure 3. Coronal holes extraction by the CATCH tool. During the eruption of the 12 July 2012 CME there were two CHs present on the Sun and both are indicated by the *green boundary* in the *top and bottom panels* of AIA 193 Å images. The *second column* shows the corresponding underlying line-of-sight magnetic field on the HMI magnetogram

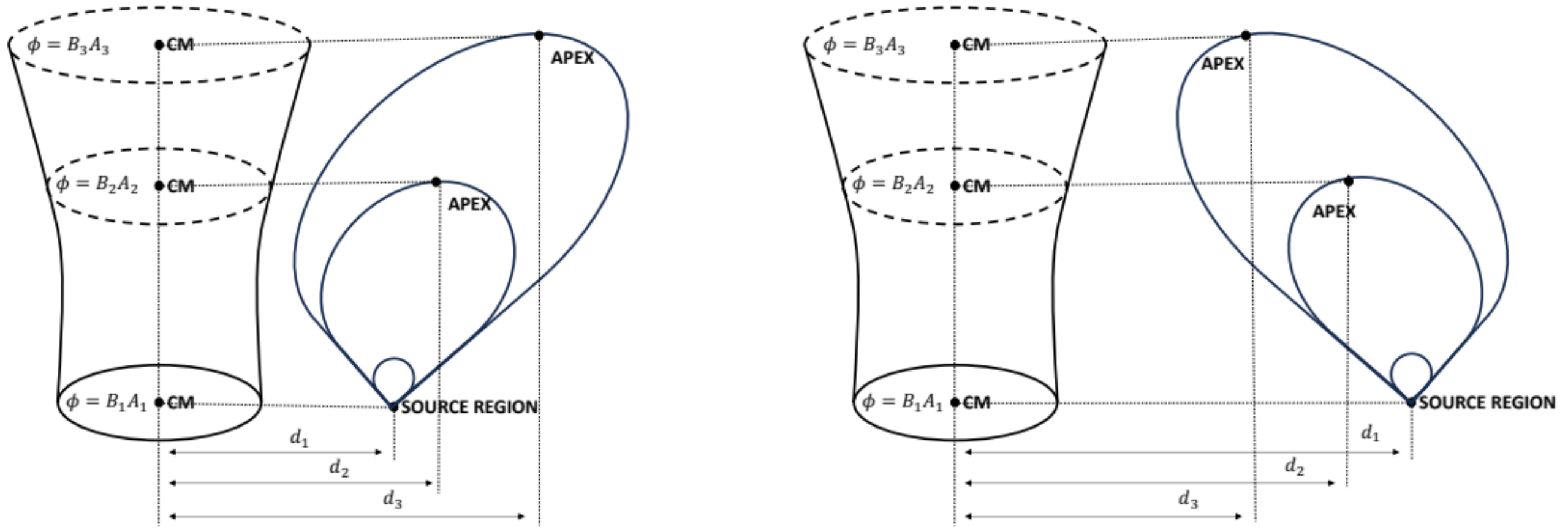

Figure 4. Change in the relative positions of the CME source location/apex projection and CH centre of mass for CME deflecting away from the CH (*left*) and CME deflecting towards CH (*right*).

Table 3. CHIP values corresponding to CHs during the eruption of the CME observed on 12 July 2012

| No. | Source | CH | Area | Mag.field | distance | CHIP |
|---|---|---|---|---|---|---|
| - | - | - | 1010 [km2] | [G] | 105 [km] | [G] |
| CH1 | W03S17 | E32S38 | 1.91 | -4.03 | 3.82 | 0.53 |
| CH2 | W03S17 | E34N25 | 0.87 | -5.18 | 4.13 | 0.26 |

Note that the only difference in calculating different CHIP values comes from the relative distance of the CME and CH. This is due to the fact that, assuming that the magnetic flux remains conserved in the CH, the product *BA* in Equation 1 remains constant. We assume that the low coronal signatures are detected at the same layer of the atmosphere where the CH reconstruction is performed and where we calculate CHIP1. In addition, we assume that the centre of mass of the coronal hole does not change with height, i.e. during the time of CME early kinematics (when we perform the GCS reconstruction). Thus, CHIP2 and CHIP3 values, which are calculated at larger heights compared to CHIP1, will change with respect to CHIP1 depending on whether the apex of the CME moves towards or away from the CH. This is sketched in Figure 4. As a consequence, for CMEs deflecting towards the CH we would expect to see decrease in relative distances and thus increase in CHIP values, and vice versa for CMEs deflecting away from CH. We note however that in case of multiple CHs, the situation might not be so simple as shown in Figure 4.

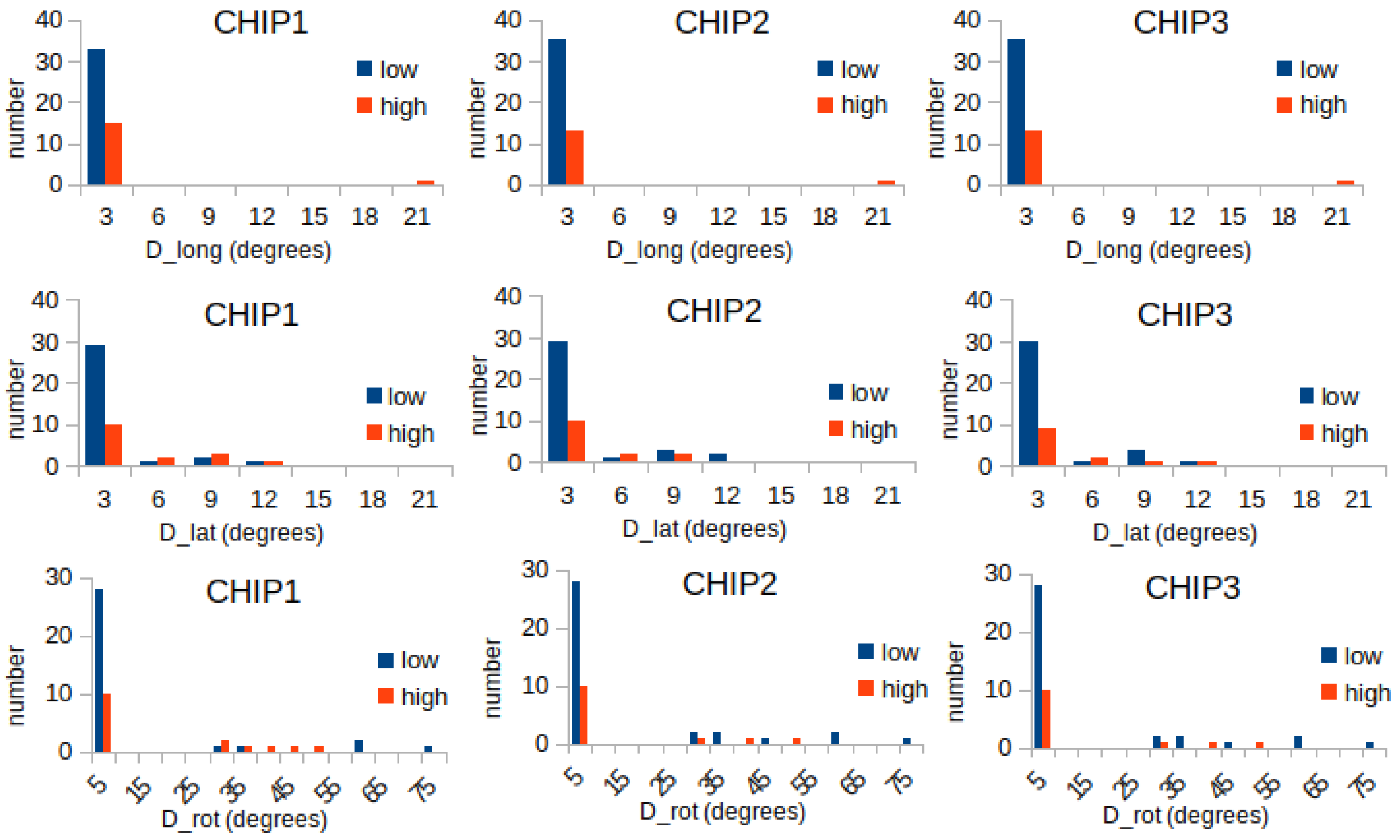

Figure 5. Distributions of changes in longitude (*upper panels*), latitude (*middle panels*), and tilt (*bottom panels*) for events with high and low CHIP value (*orange and blue, respectively*) for CHIP1 (*left panels*), CHIP2 (*middle panels*), and CHIP3 (*right panels*). The low/high CHIP values are divided according to the mean.

Results and Discussion

We first analyse the relation of the deflection/rotation of each event with respect to their total CHIP values. We consider three different CHIP values, CHIP1, CHIP2, and CHIP3 (as defined in Section 2.2), corresponding to distances obtained from the low, middle, and high corona, respectively. To test how the CMEs are affected in their deflection and rotation by the low/high CHIP values, we divide our sample according to CHIP values lower or higher than their mean value. We then check whether there is a difference between the distribution of longitude/latitude/tilt changes for low and high CHIP values. This is presented in Figure 4. We note that we also performed the analysis for the CHIP values divided into low/high CHIP values according to their median, which yielded the same results and are therefore not presented.

Figure 5 shows the distributions of changes in longitude (upper panels), latitude (middle panels), and tilt (bottom panels) for events with high and low CHIP values (orange and blue, respectively) for CHIP1 (left panels), CHIP2 (middle panels) and CHIP3 (right panels). The mean value of CHIP1/CHIP2/CHIP3 is 2.79 G / 3.12 G / 2.92 G respectively. The events are divided according to their mean CHIP1/CHIP2/CHIP3 value, as those above/below the mean are considered as low/high CHIP events. There is no obvious connection between the latitudinal deflection of CMEs and low or high CHIP1/CHIP2/CHIP3 values (middle panels in Figure 5). This suggests that the forces associated with low or high CHIP1/CHIP2/CHIP3 from the coronal holes do not impact the latitudinal deflection. Similar findings are observed for rotation (bottom-most panels in Figure 5). In longitude, only one event shows deflection towards the West which has a high CHIP1/CHIP2/CHI3 value of 4.53 G / 8.80 G / 2.06 G (upper most pannels in Figure 5). Among the 11 events that show rotation, 6 events rotate clockwise and 5 rotate anticlockwise. Out of the 10 events that show deflection in latitude, 8 events deflect toward the North and the remaining 2 towards South. More importantly, we find that 7 events deflect away from their nearby CH and 3 deflect towards it. Note that we have only one event that shows significant deflection along the longitudinal direction. The non-parametric Kolmogorov--Smirnov test shows there is no significant difference, considering the cutoff for a p-value of 1%, in the distribution means between events with low and high CHIP1/CHIP2/CHIP3 values.

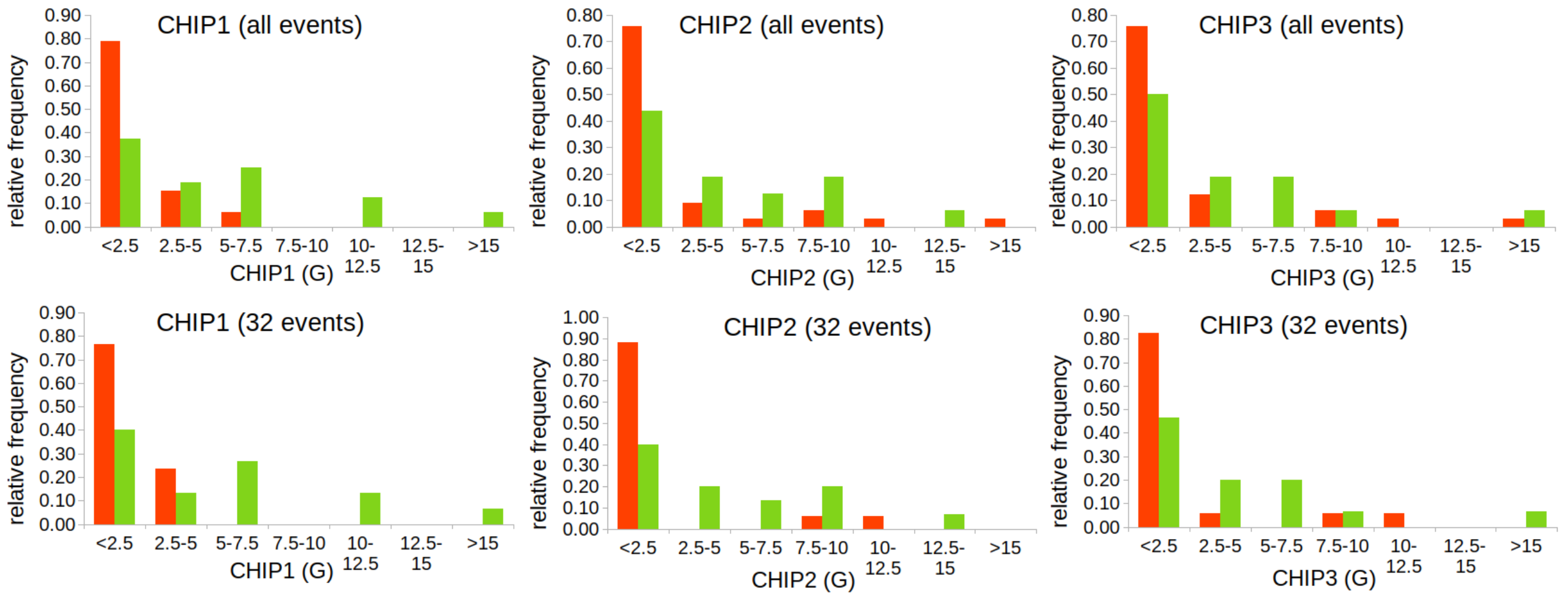

Figure 6. Distribution of CHIP1 (*left*), CHIP2 (*middle*) and CHIP3 (*right*) for events with observed change in the direction (*green*) and those without any change in the direction (*red*) for 49 events (*upper panels*) and 32 events that are tracked in both COR1 and COR2 (*lower panels*).

Considering the sketch provided in Figure 4, and the mean values of CHIP1/CHIP2/CHIP3 (2.79 G / 3.12 G / 2.92 G) our results might indicate that in general, early on CMEs deflect towards the CH (CHIP2 > CHIP1) and then away from it (CHIP3 < CHIP2).

Subsequently, we proceed to analyse the overall change in the direction of CMEs, encompassing both deflections and rotations, with respect to different CHIP values. Figure 6 displays the distribution of CHIP1 (left), CHIP2 (middle), and CHIP3 (right) for all 49 events illustrating those with observed changes in direction (green) and those without any change (red) in the upper panels. The upper panel comprises 49 events that are tracked solely in COR2 due to weak signatures in COR1 and those are tracked in both COR1 and COR2. In contrast, the lower panel includes only events that are tracked in both COR1 and COR2. We exclude 15 events tracked only in COR2, as well as 2 stealth events, where identification of the source region relied on assumptions of small brightenings and/or movement of coronal loops. In Figure 6, the upper panel reveals a significant difference between the distributions of the 49 events with and without changes in direction for CHIP1, i.e. in the lower corona (upper left panel). However, no such relationship is evident for CHIP2 and CHIP3. Statistical analysis using Student's t-test confirms the difference between events with high and low CHIP1 values, yielding a corresponding p-value of 0.0006. In the lower panels, a similar relationship is observed for the 32 events with and without changes in direction for CHIP1. Student's t-test confirms this difference, with p-values of 0.005. These findings indicate a significant difference of events with and without changes in direction for CHIP1 irrespective of whether they are observed in the COR1 and/or COR2 FOV during CME tracking. In conclusion, Figure 6 demonstrates that the influence of CHs on the direction of CMEs is significant only for CHIP1. Our results support the findings of Gopalswamy et al. (2009a), who also highlighted the role of CH in changing the direction of CMEs in the corona. Specifically, Kay et al. (2015b) found that magnetic forces dominate below 10 $R_{\odot}$, while non-magnetic forces may be responsible for interplanetary deflection/rotation.

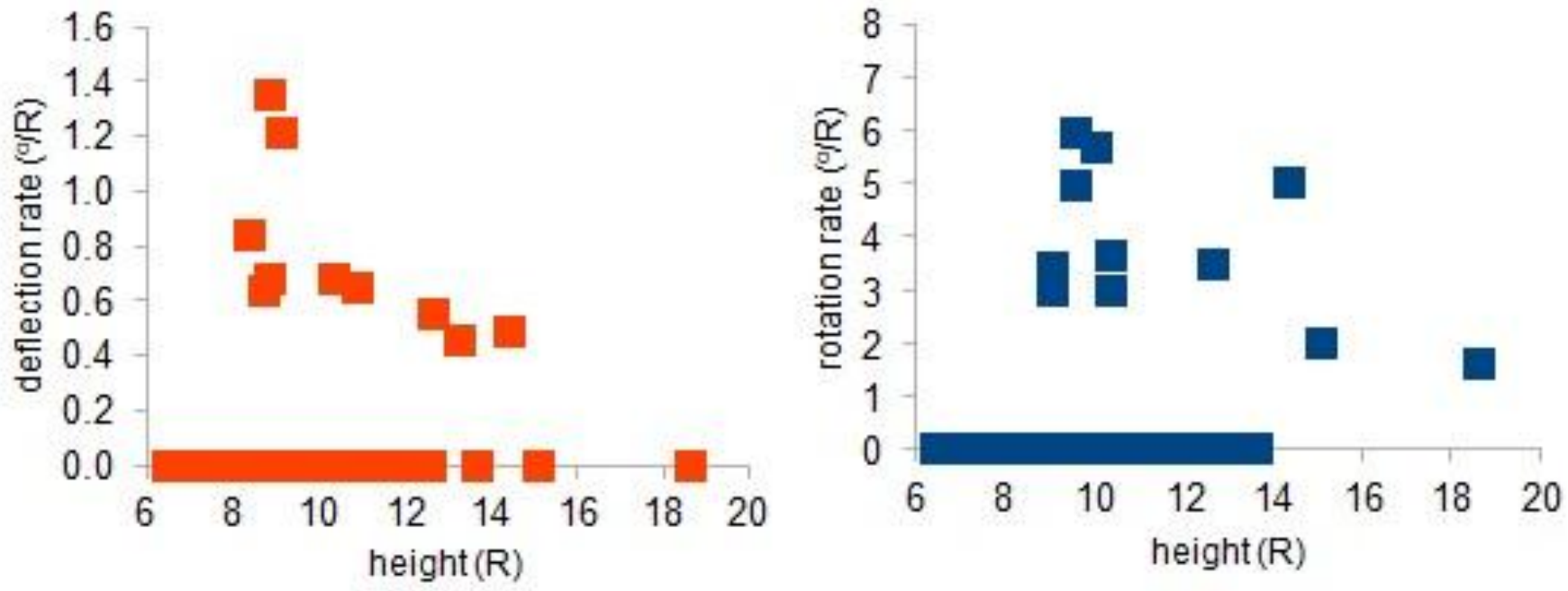

Figure 7. left panel shows the calculated deflection rate vs. height (*red*), and the *right panel* the calculated rotation rate vs. height (*blue*).

Finally, we estimate the deflection and rotation rates of CMEs to study their characteristic behaviour in the corona. We define the deflection/rotation rate as $\Delta\alpha$ / $\Delta d$, where $\Delta\alpha$ is the total deflection/rotation in longitude, latitude, and tilt, and $\Delta d$ is the change in height. Figure 7 shows the deflection and rotation rate with respect to the height (left and right). The starting height (beginning with the first GCS reconstruction) of each event is contingent upon the observable, unambiguous CME signatures in the COR1 and COR2 instruments. It is observed from the figure that the deflection rate is high below 10 $R_{\odot}$ and starts to decay beyond that. The rotation rate starts to decay beyond 15 $R_{\odot}$. Gui et al. (2011) and Sieyra et al. (2020) reported the deflection rate of CMEs in longitude and latitude over height and speed, and found that the deflection rate is higher below 4 $R_{\odot}$.

Figure 8 shows the deflection and rotation rates with respect to the CME 3D speed. The 3D speed is calculated by the linear fit to the height-time details derived from GCS reconstruction. We consider only the latitudinal and not the longitudinal deflection rate since there is only one event that deflected significantly in longitude. Both deflection and rotation rates are found to be anti-correlated with the 3D speed, which is consistent with Temmer, Preiss, and Veronig (2009) who calculated the 3D speed of 11 CMEs by applying a triangulation method and found that fast CMEs deviated less from their source region than slow CMEs.

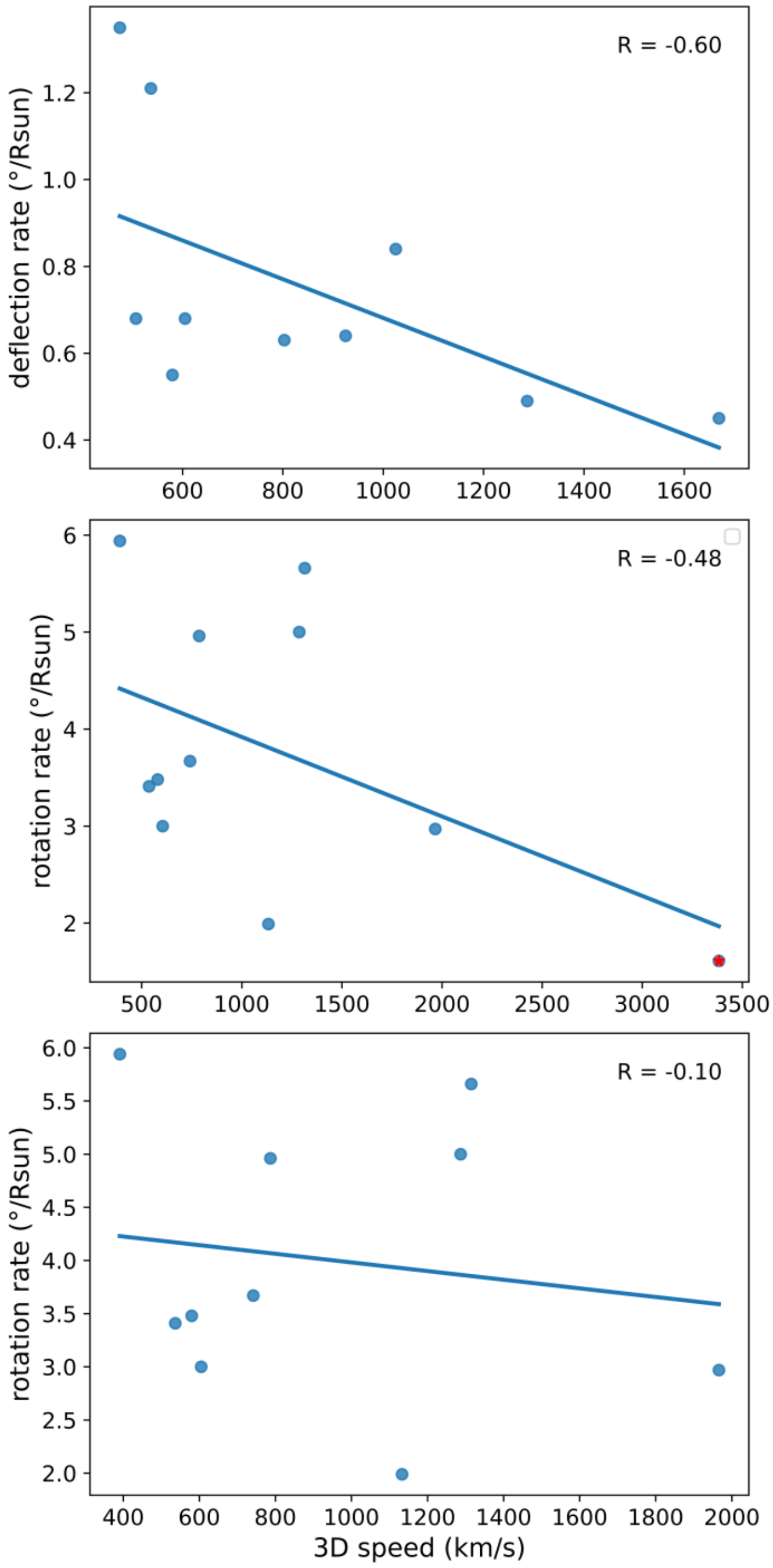

Figure 8. The upper plot shows the deflection rate vs. the 3D speed, the middle, and bottom plots show the rotation rate vs. the 3D speed with and without the outlier. The outlier is shown as a red asterisk symbol in the middle panel.

Then, we perform a bootstrapping test Efron, 1982; Efron and Tibshirani, 1991), since the number of our samples is small (10 and 11 events for deflection and rotation, respectively). This method generally regenerates a number of iterative new samples from the original sample and finds a significant correlation with the original sample. The Pearson correlation coefficient for the deflection rate and 3D speed (upper panel) is -0.60 with lower and upper confidence intervals -0.88 and 0.09 respectively. This negative correlation indicates that the deflection rate of CME decreases with the 3D speed as the distance away from the Sun. The same method is performed for the rotation rate (middle panel), and we found a correlation coefficient of -0.48 with the lower and upper confidence intervals -0.84 and 0.38 respectively (including outlier event). However, by considering the event 07 Mar 2012 as an outlier due to its relatively higher speed (3383 km $s^{-1}$), we repeat the method again for the rotation rate (lower panel) by excluding the outlier (indicated by red asterisk symbol in the middle plot) and found a correlation coefficient of -0.1. This suggests that there is no correlation found between the rotation rate and the 3D speed of events.

The studies Gopalswamy et al. (2009a) and Mohamed et al. (2012) measured the deflection of CMEs by calculating the angular distance between the Measurement Position Angle (MPA) and the Sun-Earth line. Mäkelä et al. (2013), on the other hand, estimated the deflection angle based on the angular distance between the direction obtained from the flux-rope fitting obtained from Xie, Gopalswamy, and St. Cyr (2013), and the Sun-Earth line. In contrast to these previous studies, our work utilises GCS fittings to calculate the deflection/rotation angles of CMEs and analyses their relationship with the CHIP parameter. By employing this method, we obtain a reliable measurement of the deflection/rotation in the propagation direction of CMEs. Moreover, we divide the CHIP parameter into three distinct ranges that represent different coronal regions, as explained in Section 2.2. This investigation examines the influence of CHIP on the overall change in the direction of CMEs, specifically focusing on the low, middle, and upper corona in accordance with the corresponding CHIP values. It is worth noting that CMEs typically undergo deflection or rotation relatively close to the solar surface. This suggests that the CHIP parameter plays a significant role in the overall change of CME direction in low corona.

In this article, the obtained results for the deflection and rotation of CMEs consistent with previous literature (Gui et al., 2011; Isavnin, Vourlidas, and Kilpua, 2014; Wang et al., 2014, 2016; Kay, Opher, and Evans, 2015; Kay and Opher, 2015) with the understanding that deflections below 30 $R_\odot$ are predominantly influenced by the global gradient of the magnetic field, such as the HCS and CHs, whereas the interplanetary deflections are believed to be controlled by the solar wind and interplanetary magnetic field. Based on the assumption that a large part of the deflections occur within a few solar radii from the Sun (Shen et al. (2011); Isavnin, Vourlidas, and Kilpua (2014)) due to the interactions of CH with CMEs, our analysis focused solely on the magnetic-field strength derived from CHs. We specifically examined the deflection and rotation properties of Earth-directed CMEs within the FOV of the coronagraphic observations since this interaction is less in the interplanetary medium.

## 4. Summary and Conclusions

We conducted a systematic analysis of 49 Earth-directed CMEs spanning the period from 2010 to 2020 to examine their deflection and rotation patterns. Utilising the GCS reconstruction method, we track the CMEs at various time steps and observe their behavior in the COR1 and COR2 FOV. Out of the 49 CMEs studied, one event shows significant deflection in longitude, while 10 exhibit deflection in latitude and 11 in rotation. To investigate the relationship between the CME's deflection/rotation and the influence of CHs, we calculate the influence parameter known as CHIP. This parameter is determined by considering factors such as the magnetic field, the area of the CH, and the distance between the CME source region and the CH. The CHIP parameter is divided into three categories, representing the low (CHIP1), middle (CHIP2), and upper corona (CHIP3) to examine the relationship between the influence of CHs in different regions and the propagation direction of CMEs. Analysing the deflection/rotation of CMEs separately, we found no significant deflection in latitude, longitude, and rotation between CMEs associated with low and high CHIP values in the low, middle, or upper corona. The non-parametric Kolmogorov-Smirnov test also confirms the above result. Among the 10 events exhibiting latitudinal deflection, 8 show significant deflection towards the North and 2 towards the South. Specifically, 7 events deflect away from nearby CHs, while 3 deflect towards them. The average values of CHIP1, CHIP2, and CHIP3 indicate that initially, the CMEs move towards the coronal hole and later away from it.

The deflection rate is effective within a distance of 10 $R_{\odot}$ but starts to decay beyond that, while the rotation rate begins to decline after reaching 15 $R_{\odot}$. Moreover, we observe a negative correlation between the deflection rate and the velocity of the CME, suggesting that higher velocities are associated with lower deflection rates. Furthermore, we examine the distribution of CHIP values for the CMEs with and without an overall change in their direction in the low, middle, and high corona. This analysis is based on changes in the position of the low coronal signatures and apex positions in the GCS reconstructions. In addition to the magnetic field from CHs, the velocity of the CME also plays an important role in the deflection. In contrast to previous studies, our current work employs GCS fittings to accurately calculate the deflection and rotation angles of CMEs and investigates their relationship with the CHIP parameter. By utilising this method, we achieve a reliable measurement of deflection and rotation in the propagation direction of CMEs. The CHIP categorised into three distinct categories, each representing different regions within the corona (low, middle, and high) enabled us to find a correlation between the CHIP parameter and events with or without an overall change in direction, particularly in the low corona, regardless of whether the events are tracked in both COR1 and COR2 or solely in the COR2 FOV.

This suggests a significant role played by the CHIP1 parameter in the overall change of CME direction in low corona. These outcomes contribute to a better understanding of the intricate interplay between the magnetic field derived from CHs and the deflection and rotation of CMEs.

Acknowledgements:

We acknowledge the support by the Croatian Science Foundation under the project IP-2020-02-9893 (ICOHOSS) and from the Austrian-Croatian Bilateral Scientific Project ``Multi-Wavelength Analysis of Solar Rotation Profile''. K. Martinić. acknowledges support by the Croatian Science Foundation in the scope of Young Researches Career Development Project Training New Doctoral Students.

Appendix:

Table 2.: GCS & CHIP parameters of 49 CMEs. Stealth CMEs are marked with an asterisk symbol in the first column

| Date | C2 time | Source location | Type of LCS | No. of GCS | GCS [degrees] | | | No. of CH | CHIP | Speed |
|---|---|---|---|---|---|---|---|---|---|---|
| - | [UT] | - | - | - | Δlon | Δlat | Δrot | - | [G] | [kms$^{-1}$] |
| 26 Oct 2010 | 01:36 | E00N16 | flare | 6 | 0 | -2 | -4 | 3 | 0.35 | 387 |
| 15 Feb 2011 | 02:24 | W14S13 | flare | 8 | 0 | -7 | 16 | 1 | 5.26 | 925 |
| 3 Mar 2011* | 06:12 | E09S05 | stealth | 8 | 0 | -2 | 0 | 2 | 1.79 | 329 |
| 24 Mar 2011 | 17:48 | E40S09 | flare | 4 | 0 | 0 | 0 | 2 | 2.27 | 682 |
| 25 May 2011 | 05:24 | W13S15 | flare | 4 | 0 | -3 | 0 | 3 | 5.64 | 628 |
| 2 Jun 2011 | 08:10 | E21S18 | flare | 8 | 0 | -4 | 0 | 3 | 1.52 | 1186 |
| 14 Jun 2011 | 06:12 | E31S11 | filament | 14 | 0 | 6 | 3 | 3 | 12.44 | 507 |
| 2 Aug 2011 | 06:36 | W13N10 | flare | 12 | 0 | 6 | 0 | 2 | 0.68 | 803 |
| 4 Aug 2011 | 04:12 | W36N13 | flare | 5 | 1 | 3 | -27 | 1 | 0.79 | 1966 |
| 6 Sep 2011 | 23:05 | W22N16 | flare | 6 | 0 | 0 | 0 | 1 | 0.79 | 925 |
| 13 Sep 2011 | 22:10 | W13N15 | flare | 10 | 0 | 0 | 0 | 2 | 1.92 | 699 |
| 19 Sep 2011 | 06:00 | E62N16 | flare | 6 | 0 | 0 | 0 | 1 | 0.93 | 920 |
| 24 Sep 2011 | 12:48 | E58N12 | flare | 7 | 0 | 0 | 0 | 1 | 1.09 | 1524 |
| 2 Oct 2011 | 02:00 | W13N03 | flare | 11 | 0 | 0 | 0 | 1 | 0.4 | 576 |
| 22 Oct 2011 | 01:25 | W38N25 | filament | 7 | 0 | 0 | -18 | 1 | 1.44 | 659 |
| 27 Oct 2011 | 12:00 | E17N29 | filament | 11 | 0 | 0 | 23 | 2 | 2 | 632 |
| 9 Nov 2011 | 13:36 | E43N19 | filament | 7 | 0 | 0 | -57 | 1 | 0.1 | 1315 |
| 26 Dec 2011 | 11:48 | W02N21 | filament | 12 | 0 | 1 | 48 | 1 | 5.49 | 787 |
| 18 Jan 2012 | 12:24 | E03S03 | filament | 14 | 0 | 0 | 57 | 1 | 0.21 | 391 |
| 19 Jan 2012 | 14:36 | E25N46 | flare | 12 | -9 | 0 | 0 | 1 | 0.23 | 1047 |
| 24 Feb 2012 | 03:46 | E23N17 | filament | 9 | 0 | 2 | -38 | 4 | 5.86 | 742 |
| 7 Mar 2012 | 00:24 | E30N24 | flare | 5 | 0 | 0 | -30 | 2 | 3.52 | 3383 |
| 13 Mar 2012 | 17:36 | W67N21 | flare | 7 | 0 | 0 | 0 | 1 | 0.49 | 2148 |
| 12 May 2012 | 00:00 | E12S12 | filament | 8 | 0 | -6 | 0 | 2 | 6.85 | 1669 |
| 12 Jul 2012 | 16:48 | W03S17 | flare | 12 | 0 | -7 | 72 | 2 | 0.46 | 1287 |
| 28 Sep 2012 | 00:12 | W30N03 | filament | 9 | 0 | 2 | -2 | 1 | 0.05 | 1175 |
| 5 Oct 2012* | 02:48 | W04S24 | stealth | 9 | 0 | -11 | 31 | 2 | 3.74 | 537 |
| 27 Oct 2012 | 16:48 | W32N03 | filament | 8 | 0 | 4 | 0 | 2 | 0.7 | 287 |
| 9 Nov 2012 | 15:12 | E15S19 | filament | 12 | 0 | -7 | -44 | 1 | 11.07 | 580 |
| 23 Nov 2012 | 13:48 | E13S40 | filament | 10 | -1 | -7 | 31 | 2 | 1.76 | 605 |
| 13 Jan 2013 | 12:00 | E09N13 | flare | 7 | 0 | 0 | 0 | 3 | 1.79 | 373 |
| 15 Mar 2013 | 07:12 | E10N19 | flare | 11 | -1 | -3 | -30 | 3 | 15.88 | 1133 |
| 9 Jul 2013 | 15:12 | E25N20 | filament | 12 | 0 | 2 | 0 | 2 | 0.02 | 771 |
| 29 Sep 2013 | 22:12 | W29N14 | filament | 8 | 0 | 0 | 0 | 2 | 3.17 | 1181 |
| 12 Dec 2013 | 03:36 | W51S31 | flare | 5 | 0 | 0 | 0 | 1 | 1.44 | 1040 |
| 4 Feb 2014 | 01:25 | W05S08 | flare | 4 | 0 | 0 | 0 | 2 | 1.62 | 669 |
| 12 Feb 2014 | 06:00 | W00S06 | flare | 5 | 0 | 0 | 0 | 2 | 1.69 | 648 |
| 2 Apr 2014 | 13:36 | E56N17 | flare | 6 | 0 | 0 | 0 | 2 | 0.33 | 1535 |
| 18 Apr 2014 | 13:25 | W33S14 | flare | 5 | 7 | 0 | 0 | 4 | 5.7 | 1335 |
| 4 Jun 2014 | 15:24 | E40S24 | filament | 9 | -1 | -12 | 5 | 2 | 1.54 | 475 |
| 15 Aug 2014 | 17:48 | W07S17 | filament | 6 | 0 | 4 | 0 | 1 | 1.31 | 662 |
| 10 Sep 2014 | 18:00 | E06N07 | flare | 4 | 0 | 0 | 0 | 2 | 2.78 | 1310 |
| 28 Dec 2015 | 12:12 | W23S11 | flare | 6 | 20 | -7 | 22 | 3 | 4.53 | 1025 |
| 5 Nov 2016 | 04:24 | W10N40 | filament | 6 | 0 | 3 | 0 | 4 | 3.36 | 551 |
| 23 May 2017* | 05:00 | W18S09 | stealth | 4 | 0 | 0 | 0 | 6 | 1.74 | 410 |
| 14 Jul 2017 | 01:25 | W32S10 | flare | 6 | 0 | 0 | 0 | 4 | 0.89 | 1562 |
| 4 Sep 2017 | 20:36 | W13S19 | flare | 4 | 0 | 0 | 0 | 2 | 2.22 | 1650 |
| 6 Sep 2017 | 12:24 | W39S14 | flare | 7 | 0 | 0 | 0 | 3 | 2.97 | 1587 |
| 12 May 2019 | 20:24 | W04N02 | filament | 10 | 0 | 0 | 0 | 2 | 3.87 | 529 |